\documentclass[a4paper,10pt]{article}
\usepackage[utf8]{inputenc}

\title{   Black Holes Thermodynamics in a new kind of Noncommutative Geometry }
\author{Mir Faizal$^1$, R. G. G. Amorim$^{2,3}$, S. C. Ulhoa$^{2,4}$ \\\\
$^1$Department of Physics and Astronomy, University of Waterloo, \\Waterloo,
Ontario, N2L 3G1, Canada.\\
$^2$Instituto de F\'{i}sica,
Universidade de Bras\'{i}lia, 70910-900, \\ Bras\'{i}lia, DF,
Brazil.\\ 
$^3$Faculdade Gama, Universidade de Bras\'{i}lia, \\ Setor Leste
(Gama), 72444-240, \\ Bras\'{i}lia-DF, Brazil.\\
$^4$International Center of Condensed Matter Physics,\\
Universidade de Bras\'{i}lia,\\  70910-900, Bras\'{i}lia, DF,
Brazil.}
\date{}
\begin{document}

\maketitle

\begin{abstract}
Motivated by the energy dependent metric in gravity's rainbow, we will propose a new kind of energy dependent 
noncommutative geometry. It will be demonstrated that like gravity's rainbow, this new noncommutative geometry 
is described by an energy dependent  metric. We will analyse the effect of this noncommutative deformation on 
 the Schwarzschild black holes and Kerr black holes. 
We will perform our analysis by relating the commutative and this new energy dependent 
noncommutative metrics using an energy dependent Moyal star product.
We will also analyze the thermodynamics of these new noncommutative black hole solutions. We will explicitly
derive expression for the corrected entropy and temperature  for these black hole solutions.
It will be demonstrated  that for  these deformed solutions  
    black remnants cannot form. This is because these correction increase rather than reduce the temperature 
   of the black holes.
   \end{abstract}

\section{Introduction}

The renormalization group flow predicts that the constants in a quantum field theory will flow and depend on the scale
at which such a theory is probed  \cite{pppp2}.  However, the scale at which a theory is probed will in turn depend on the energy of 
the probe used in that theory. Hence, we can argue that   the coupling constants depend on the energy of the probe. In fact, such 
scale dependence (energy dependence)  of constants has been used for analyzing the   flow of the cosmological constant \cite{pppp2a} and the Newton constant \cite{pppp4}. 
It has thus been demonstrated using renormalization group flow that these constants depend on  the scale at which a theory is probed, 
and so they depend on the energy of the probe.
It may also be noted that   string theory can be viewed as a two dimensional conformal field theory, and in this formalism the  target 
space metric can be viewed as a matrix of coupling constants for this two dimensional conformal field theory. 
Hence, it is expected that this matrix of coupling constants will also flow, make the spacetime metric scale dependent. However, as 
the scale at which spacetime is probed will depend on the energy of the probe, it is expected that the spacetime metric will also become
energy dependent. It may be noted that this energy dependent metric is the basis of gravity's rainbow  \cite{gr, rg, r1, r2}. 
As it is expected that spacetime
geometry will be an emergent structure, it is not that surprising that 
the geometry will depend on the energy of the probe. Furthermore, in almost all theories of quantum gravity, for example, the discrete spacetime
\cite{1g}, models based on string field theory \cite{2g}, spacetime
foam \cite{4g}, spin-network in loop quantum gravity (LQG)
\cite{5g},   noncommutative geometry 
\cite{Carroll,FaizalMPLA},  and ghost condensation \cite{FaizalJPA}, the spacetime geometry  is expected 
to break down at Planck scale, (hence at Planck energy). This can only occur if we have some dependence of scale in the metric, and this  is achieved
in gravity's rainbow by making the metric depend explicitly on the energy of the probe 
(and hence implicitly on the scale at which spacetime is probed). 

Finally, we would like to comment that in gravity's rainbow,  this energy (scale) dependence of the metric will only
become effective at Planck energy, 
and in the IR limit this theory coincides with  general relativity. Furthermore, this depends only  on the energy of the probe,
which fixes the scale at which the spacetime is probed, and not on the kind of particle used to probe the spacetime.
It may be noted that constraints
for such energy dependence of the metric from observational data has also been studied \cite{pppp7}. 
It has been argued that the energy of a particle emitted
in Hawking radiation, near the horizon can be used to fix the scale at which the black hole is being probed, as this particle can act as an effective probe
for the geometry \cite{5A}. This can be used to incorporate the energy dependence in a black hole metric. 
It has also been demonstrated that this energy dependence of metric   can alter the physics at the last stages of the evaporation of black holes 
 \cite{2A, 4A}. This can also  have important phenomenological consequences for the detection of mini black holes at particle collides \cite{6B}.
In the rainbow gravity, the  energy dependence of the metric is usually incorporated through rainbow functions. 

It may be noted that  just like the   Horava-Lifshitz gravity \cite{HoravaPRD, HoravaPRL}, gravity's rainbow \cite{gr, rg, r1, r2}
has also been proposed as a UV completion of general 
relativity. 
In the  Horava-Lifshitz gravity \cite{HoravaPRD, HoravaPRL}, 
the scaling properties of space and time 
are changed, and in the  gravity's rainbow the metric of the spacetime becomes scale dependent. This  scale dependence of the metric 
is incorporated by making the metric depend on  the energy of the probe which probes the geometry of spacetime. In fact, it 
  has   been demonstrated that the gravity's rainbow and Horava-Lifshitz gravity are actually related to each other 
for specific choices of rainbow functions \cite{re}.  The gravity's rainbow has also been studied using 
  Finsler geometries   \cite{fg12, fg14, fg16,fg18, fg19}.
Thus, there is a strong physical motivation to make the geometry dependent on the energy of the probe (and thus the scale at which 
the theory is probed). 
So, 
in this paper, we will propose an alternative way to incorporate this energy dependence in the geometry. We will propose that the parameter used
to define the noncommutative deformation depends on the energy of the probe (hence the scale at which the spacetime is probed). In fact, 
as string theory 
can be used  to motivate both  the energy dependence of the emergent spacetime geometry   and noncommutativity,  it is possible to relate these two
effects together. Furthermore, as the thermodynamics of rainbow black holes has been thoroughly studied \cite{pppp8a,pppp8b,pppp8c,pppp8d,ga, 1A, q5}, 
we will analyse the effect of this
new proposal on the thermodynamics of black holes. 
In this structure, the noncommutativity will depend on the energy of the test particle. Hence, just like gravity acts differently 
on particles of different energies in gravity's rainbow, the deformation of gravity will be different for particles of different energies 
in this new theory of noncommutativity. 

In this paper, we will analyse the effect of this new deformation on the thermodynamics of black holes. It may be noted that 
in the original proposal of black hole thermodynamics, the entropy scaled with the area of the horizon \cite{1, 1a, 2, 4, 4a}.
This observation
  has led to the
development of the holographic principle~\cite{5, 5a}.
In fact, the precise form of the area-entropy relation for black holes can be written as
  $S = A/4$,
where $S$ is the entropy of the black hole and $A$ is the area.
However, this area-entropy relation is expected to get corrected as the black holes become small 
in size due to Hawking radiation. This is because     the
quantum fluctuations  become important at that scale, and the quantum corrections in turn correct the black hole entropy. 
So, at small scales it is expected that
the black hole thermodynamics will get modified, and this is also expected to modify the
 holographic
principle \cite{6, 6a}.
It may be noted that such corrections have been derived using various approaches, such as the
  non-perturbative quantum   general
relativity  approach   \cite{1z}, the  Cardy formula
\cite{card, ca}, the partition function  has been computed for
BTZ black holes \cite{gks}, and
  string theoretical effects  \cite{solo1, solo2, solo4, solo5, jy}.
 The  noncommutativity also produces  corrections terms   in the black hole thermodynamic
 \cite{a,b,c,d}. The effect of noncommutativity is usually studied by    replacing
  point-like structures in the original theory by   objects which are smeared in
  spacetime. This is done by replacing the delta function by  a
  Gaussian distribution with minimal width.
  The width of this Gaussian distribution is fixed by
    the noncommutative parameter. In this paper, we will analyse the black holes in this new noncommutative theory, which 
    is  motivated by gravity's rainbow. This will be done by following a procedure similar to the 
    one used for analysing black hole thermodynamics in gravity's rainbow \cite{5A}. Thus, we will first fix an energy scale 
from the  energy of a particle near the horizon, and then use this energy scale to analyse the effects of this new noncommutativity on 
the thermodynamics of black holes. 

\section{Energy Dependent Noncommutative Geometry}
 In this section, we will review the construction of noncommutative gravity \cite{moff,moff1, az12,W}.
 In the  noncommutative geometry, spacetime coordinates are promoted to
 a set of noncommutative self-adjoint operators, such that they
 satisfy
 \begin{equation}
  [x^\mu,  x^\nu] = i \theta^{\mu\nu},
 \end{equation}
where $\theta^{\mu\nu}$ is a  antisymmetric matrix.
 The product of two fields on this noncommutative spacetime
 can be replaced by a Moyal product of commutative fields, where
 the   Moyal product  is given by \cite{Moyal}
\begin{eqnarray}
f(x)* g(x)
&=& \exp\biggl[\frac{i}{2}\frac{\partial}{\partial
x^\mu}\theta^{\mu\nu}\frac{\partial}{\partial
y^\nu}\biggr]f(x)g(y)\mid_{y\rightarrow x}
\nonumber \\ &=& f(x)g(x)+\frac{i}{2}\theta^{\mu\nu}\partial_\mu f(x)\partial_\nu
g(x)+{\cal O}(\theta^2).
\end{eqnarray}
It may be noted that
the noncommutative
field theories are nonlocal as the Moyal product of fields involves
  an infinite number
of derivatives. Furthermore,    as the
  spacetime coordinates do not commute, the  noncommutativity gives rise to
 a minimum length in spacetime
$\Delta x^\mu\Delta x^\nu \geq \frac{1}{2}\vert\theta^{\mu\nu}\vert
$.

  We can employ the Weyl quantization
procedure~\cite{Wess,Weyl} to relate  the metric tensor on noncommutative
spacetime ${\hat g}_{\mu\nu}$ with the metric on commutative spacetime
$g_{\mu\nu}$.  Now the   Fourier transform of the metric is given by
\begin{equation} {\tilde g}_{\mu\nu}(k)=\frac{1}{(2\pi)^2}\int
d^4x \exp(-ik_\sigma x^\sigma)g_{\mu\nu}(x), \end{equation}
so, we can write
\begin{equation}
{\hat g}_{\mu\nu}(\hat x)=\frac{1}{(2\pi)^2}\int d^4k\exp(ik_\sigma{\hat
x}^\sigma){\tilde g}_{\mu\nu}(k).
\end{equation}
Thus, the operators ${\hat g}_{\mu\nu}$ and ${\hat x}$ replace the variables
$g_{\mu\nu}$ and $x$.
The product of two tensor fields on this noncommutative spacetime
is equal to the Moyal product of two  tensor fields on ordinary spacetime. So, the product of ${\hat
f_{\mu\nu}}$ and ${\hat g_{\mu\nu}}$ is equal to $f(x)_{\lambda \sigma }* g(x)_{\tau\rho}$, such that   \cite{moff, moff1},
\begin{eqnarray}
f(x)_{\lambda \sigma }* g(x)_{\tau\rho}&=&
\frac{1}{(2\pi)^4}\int
d^4kd^4p\exp[i(k_\mu+p_\mu)x^\mu-\frac{i}{2}k_\mu\theta^{\mu\nu}p_\nu] \nonumber \\ && \times
{\tilde
f_{\lambda \sigma }}(k){\tilde g_{\tau\rho}}(p)
\nonumber \\ &=&
 \exp\biggl[\frac{i}{2}\frac{\partial}{\partial
x^\mu}\theta^{\mu\nu}\frac{\partial}{\partial
y^\nu}\biggr]f_{\lambda \sigma }(x)g_{\tau\rho}(y)\mid_{y\rightarrow x},
\end{eqnarray}
where ${\tilde f_{\lambda \sigma }}(k)$ is the Fourier transform
\begin{equation}
{\tilde f_{\lambda \sigma }}(k)
=\frac{1}{(2\pi)^2}\int d^4x\exp(-ik_\sigma x^\sigma)f_{\lambda \sigma }(x).
\end{equation}
In presence of matter the   situation with regards to noncommutativity of
spacetime coordinates can be imposed on the
 vierbeins as
\begin{equation}
g_{\mu\nu}=e^a \,_{(\mu}  * e^b \,_{\nu)} \eta_{ab}.
\end{equation}

Now motivated by gravity's rainbow, we will define a new energy dependent noncommutative geometry. We will also 
demonstrate that just like the gravity's rainbow, it will also produce an energy dependent metric. 
In gravity's rainbow, a one-parameter family of 
energy-dependent orthonormal frame fields give rise to a one-parameter family of energy-dependent metrics \cite{gr, rg, r1, r2}
\begin{equation}
g^{\mu\nu}(E/E_{p})=e_{a}^{\mu}(E/E_{p})e^{\nu}_b(E/E_{p})\eta^{ab},  
\end{equation}
where $E$ is the energy of the test particle, and $E_P$ is the Planck energy. 
So,  motivated by gravity's rainbow, let us assume that the commutativity depends on the energy of the test particle, 
 \begin{equation}
  [x^\mu,  x^\nu] = i \theta^{\mu\nu} (E/E_P),
 \end{equation}
where we require 
\begin{equation}
 \lim_{E \to 0} \theta^{\mu\nu} (E/E_P) \to 0. 
\end{equation}
It may be noted that various different choices for $\theta^{\mu\nu}$ have been studied \cite{u1, u2, u0, u4}.
The noncommutativity is expected to arise because of background fluxes in string theory \cite{ncb1,ncb2}. 
It has also been demonstrated that certain background field can break the Lorentz symmetry of the system \cite{l1, l0}. 
Furthermore, the Lorentz symmetry breaking has been studied for noncommutative field theories \cite{b1, b2, b0, b4}. 
It is the breaking of Lorentz symmetry in the discrete spacetime
\cite{1}, models based on string field theory \cite{2}, spacetime
foam \cite{4}, spin-network in loop quantum gravity (LQG)
\cite{5},   noncommutative geometry 
\cite{Carroll,FaizalMPLA},  and ghost condensation \cite{FaizalJPA},  which has been used as a motivation to construct 
gravity's rainbow. 
Thus, motivated by this, we will now  consider the
  case where the non-vanishing
component of $\theta^{\mu\nu}(E/E_P)$ are $\theta^{23}(E/E_P)$ and $\theta^{32} (E/E_P)$.
This would break the Lorentz symmetry in case of flat spacetime. We will now analyse the effect of such 
a noncommutative parameter on a  Schwarzschild solution. 
So, we choose the following value for the noncommutative coordinate
\begin{equation}
\theta^{\mu\nu}=  \left[ \begin {array}{cccc}
0&0&0&0\\\noalign{\medskip}0&0&0&0
\\\noalign{\medskip}0&0&0&\beta(E/E_p)\\\noalign{\medskip}0&0&-\beta(E/E_P)&0
\end {array} \right], \label{ab}
\end{equation}
where we take the following simple energy dependent form for $\beta(E/E_P)$ 
\begin{equation}
 \beta(E/E_P) = \gamma\left(\frac{E}{E_P (E- E_P)}\right)^\eta, 
\end{equation}
where $\eta$ and $\gamma$ are parameters in the theory, which can be determined from experiments. 
It may be noted that 
\begin{equation}
 \lim_{E \to 0} \beta (E/E_P) \to 0. 
\end{equation} 
Thus, in the IR limit, we obtain the usual commutative theory, and only at very large energies, do we observe 
noncommutativity. 
Furthermore, the structure of spacetime breaks down at $E_P$, as 
\begin{equation}
 \lim_{E \to E_P} \beta (E/E_P) \to \infty. 
\end{equation}
Now $E_P$ is the maximum energy that any object can attain. The existence of a maximum energy scale has also 
been used as a motivation for the development of gravity's rainbow \cite{gr, rg, r1, r2}. The energy dependence 
is incorporated in gravity's rainbow using rainbow functions. Various rainbow functions have been proposed using 
different theoretical and experimental considerations  \cite{4g, pppp7, rf12a, rf12b}. Using these rainbow functions 
different form of energy dependence of different geometry  have been studied
\cite{5A, 6B, 2A, ga}. 
We have now observed that such a
maximum energy scale can also be incorporated into this new kind of noncommutative geometry. 
However, we would like to point out that we have only used a very simple form of energy dependence, 
such that it statisfied the essential properties of such an energy dependent geometry.  
It is important to investigate other forms of energy dependent
noncommutativity, and analyze its relation 
with other theoretical and experimental approaches. 
It would also be interesting to 
investigate if bounds for  parameter $\eta$ and $\gamma$ 
can be obtained from different consideration similar to the bounds obtained for rainbow functions \cite{pppp7}.
However, in this paper, we will restrict our analysis to this simple form of energy dependence. We can also 
consider the simple case, when $\eta \sim \gamma \sim 1 $, and so $ \beta(E/E_P) \sim {E}/ E_P (E- E_P) $. 

Now we can define the Moyal product for such
an energy dependent noncommutativity as
\begin{eqnarray}
f(x)_{\lambda \sigma }[\otimes_{(E/E_P)}] g(x)_{\tau\rho} &=&
 \exp\biggl[\frac{i}{2}\frac{\partial}{\partial
x^\mu}\theta^{\mu\nu} (E/E_P)\frac{\partial}{\partial
y^\nu}\biggr]\nonumber \\ && \times 
f_{\lambda \sigma }(x)g_{\tau\rho}(y)\mid_{y\rightarrow x}.
\end{eqnarray}
We can now also write the metric as 
\begin{equation}
g_{\mu\nu}=e^a \,_{(\mu} [\otimes_{(E/E_P)}] e^b \,_{\nu)} \eta_{ab}.
\end{equation}
It may be noted that just like gravity's rainbow, this metric is energy dependent. 

The spin connection for this noncommutative spacetime can be written as 
\begin{eqnarray}
\omega_\mu^{ab}&=& 2e^{\nu a}[\otimes_{(E/E_P)} ] \partial_\mu   e_\nu^b-2e^{\nu b} [\otimes_{(E/E_P)}] \partial_\mu e_\nu^a\nonumber \\ &&
-2e^{\nu a}[\otimes_{(E/E_P)}] \partial_\nu e_\mu^b+2e^{\nu b}[\otimes_{(E/E_P)}]  \partial_\nu e_\mu^a
\nonumber \\ && +e_{\mu c}[ \otimes_{(E/E_P)}] e^{\nu a}[\otimes_{(E/E_P)}]e^{\sigma b}[\otimes_{(E/E_P)}] \partial_\sigma e_\nu^c\nonumber \\ && 
-e_{\mu c} [
\otimes_{(E/E_P)}] e^{\nu a}[\otimes_{(E/E_P)} ] e^{\sigma b}[\otimes_{(E/E_P)}]\partial_\nu e_\sigma^c.
\end{eqnarray}
The spin connection is subject to the gauge
transformation
\begin{equation}
(\omega_\mu)^a_b\rightarrow[U_C[\otimes_{(E/E_P)}]\omega_\mu [\otimes_{(E/E_P)}]
U_C^{-1}-(\partial_\mu U_C)[\otimes_{(E/E_P)}] U_C^{-1}]^a_b,
\end{equation}
where $U_C$ is an element of a  noncommutative group of
orthogonal transformations $NCSO(3,1)$ \cite{moff1}.
The noncommutative curvature tensor can be written as \cite{moff} 
\begin{eqnarray}
(R_{\mu\nu})^a_b&=& \partial_\mu(\omega_\nu)^a_b-\partial_\nu(\omega_\mu)^a_b
+(\omega_\mu)^a_c[\otimes_{(E/E_P)}](\omega_\nu)^c_b\nonumber \\ && -(\omega_\nu)^a_c[\otimes_{(E/E_P)}]
(\omega_\mu)^c_b,
\end{eqnarray}
This noncommutative curvature tensor transforms as
\begin{equation}
(R_{\mu\nu})^a_b\rightarrow
U^a_{Cc}[\otimes_{(E/E_P)}](R_{\mu\nu})^c_d[\otimes_{(E/E_P)}](U_C^{-1})^d_b.
\end{equation}
The   action for Einstein gravity deformed by noncommutativity is given by \cite{moff,moff1}
\begin{eqnarray}
S &=& \frac{1}{2\kappa}\int d^4x[e[\otimes_{(E/E_P)}]e^\mu_a[\otimes_{(E/E_P)}]
(R_{\mu\nu})^a_b[\otimes_{(E/E_P)}] e^{\nu b}] \nonumber \\ &&  + \int d^4x[e[\otimes_{(E/E_P)}] \mathcal{L}_{matter}]
\end{eqnarray}
where $e=\sqrt{-g}$. So, we can write the noncommutative  Einstein equation as
\begin{eqnarray}
 R^a_\mu - \frac{1}{2} R[\otimes_{(E/E_P)}]e^a_\mu &=& -\kappa T^a_\mu,
\end{eqnarray}
where 
\begin{eqnarray}
 R^b_\mu &=&(R_{\mu\nu})^{ab} [\otimes_{(E/E_P)}]e^{\nu}_a, \nonumber \\
 R &=& R^a_\mu[\otimes_{(E/E_P)}]e_a^\mu, 
\end{eqnarray}
and  $T^a_\mu$ is the 
energy momentum tensor for matter fields in this energy dependent  noncommutative spacetime. 
It may be noted that this noncommutative metric  induces a deformed diffeomorphism group. There exists a map between
this deformed diffeomorphism group  and the original
diffeomorphism group \cite{nonab}. 
It is possible to obtain the solutions to the noncommutative Einstein equation by performing a noncommutative deformation of  the solutions to 
usual  Einstein equation \cite{moff, moff1}.

\section{New Noncommutativity for  Black Holes}\label{spherical}
In this section, we will analyse the noncommutative  black hole solutions.
This will be obtained by deforming 
the usual black hole solutions.  The uncertainty principle, $\Delta p \geq 1/ \Delta x$, can be used to obtain a lower bound on the energy,
$\Delta E \geq 1/ \Delta x$ of a particle in the Hawking radiation, which is near the horizon, and is used as a probe for the horizon. 
The value of the uncertainty in position can be taken to be the event horizon radius, 
$ E \geq 1/\Delta x \sim {1}/{r_+}, 
$ where $r_+$ is the radius of the horizon. This has been used to modify the thermodynamics of black holes in gravity's rainbow \cite{5A, 6B, 2A, ga}. 
We shall use this energy to set the scale for the energy $E$ used in noncommutativity $\beta (E/E_P)$. 
Now we will deform a Schwarzschild solution. 
We can write the original metric for a Schwarzschild solution as
\begin{equation}
ds^2=-\left(1-\frac{2M}{r}\right)dt^2+\left(1-\frac{2M}{r}\right)^{-1}dr^2
+r^2\left(d\theta^2+\sin^2{\theta}d\phi^2\right)\,,
\end{equation}
where $M$ is the mass of the black hole. Furthermore, we can write the
 tetrad field for this system, and deform it by the energy dependent noncommutativity given by Eq. (\ref{ab}), and obtain 
\begin{equation}
\sqrt{-\tilde{g}}=\sqrt{-g}+\beta^2(E/E_p)\sqrt{\frac{-g_{00}g_{11}g_{22}}{16g_{33}}}
\frac{\partial^2 g_{33}}{\partial\theta^2}.
\end{equation}
Now we can calculate the entropy of the noncommutative Schwarzschild black hole and its temperature.
The entropy of the noncommutative Schwarzschild black hole is given by
\begin{eqnarray}
\tilde{S}(r_+)&=&\left(1-\frac{\beta^2(E/E_p)}{4}\right)S(r_+)\nonumber \\
&=& \left(1-\frac{\gamma^2}{4}\left(\frac{E^2}{ E_P^2 (E- E_P)^2}\right)^\eta \right)S(r_+)
\end{eqnarray}
where $S_+ = A/4$ is the entropy of the original commutative Schwarzschild black hole,
and $\tilde{S}(r_+)$ is the entropy of the noncommutative  Schwarzschild black hole.
The temperature is defined as $T^{-1} = \partial \tilde{S}/ \partial M$, and so 
 we can write the temperature of the  noncommutative Schwarzschild black hole as
\begin{eqnarray}
 T &=& \frac{1}{8\pi M} \left(1+\frac{\beta^2(E/E_p)}{4}\right)\nonumber \\ &=&
 \,\frac{1}{8\pi M} \left(1+ \frac{\gamma^2}{4} \left(\frac{E^2}{ E_P^2 (E- E_P)^2}\right)^\eta\right).
\end{eqnarray}
 As we can see the noncommutativity modifies the thermodynamics of the original
 Schwarzschild black hole.  Here we have first obtained the corrections to the entropy of the  Schwarzschild black hole, 
 and then used these corrections to obtain the corrections to the temperature of the  Schwarzschild black hole. 
It may be noted that for this deformed solution, 
   a black remnant cannot form,  as  these correction increase rather than reduce the temperature 
   of the black hole. 
We would also like to point out that   black remnants form   in gravity's rainbow only if an undeformed dispersion relation
is used  \cite{2A, 4A, 6B}, and  black hole remnants do not form if  the modified dispersion relation are used  \cite{f15a}.
 However, this argument cannot be used in this paper, as this paper only uses 
 perturbative calculations. 
 
Now we will study the thermodynamics of a noncommutative Kerr black hole. The
 general form of the line element of a spacetime with    axial symmetry can be written as
\begin{equation}
ds^2=
g_{00}dt^2+2g_{03}\,d\phi\,dt
+g_{11}dr^2 +g_{22}d\theta^2+ g_{33}d\phi^2\,,
\label{3.1}
\end{equation}
where the metric components are functions of $r$ and $\theta$.
Now we can write the    tetrad field for this system, and again  the energy dependent noncommutativity given by Eq. (\ref{ab}).  
It will be useful to define, $\Delta g_{\mu\nu}=\tilde{g}_{\mu\nu}-g_{{\mu\nu}}$, and 
  using this definition, we
can write
\begin{equation}
\tilde{S}=\frac{1}{4}\int\int\sqrt{-g}\left[1+\frac{1}{2}\left(\frac{\Delta g_{11}}{g_{11}}+\frac{\Delta g_{22}}{g_{22}}\right)-\frac{g_{00}\Delta g_{33}}{2\delta}\right]d\theta d\phi
\end{equation}
The metric for the
Kerr black hole can be written as
\begin{equation}
ds^2=-\frac{\psi^2}{\rho^2}dt^2-2\frac{\chi\sin^2\theta}{\rho^2}dtd\phi+\frac{\rho^2}{\Delta}dr^2+\rho^2d\theta^2+\frac{\Sigma^2\sin^2\theta}{\rho^2}d\phi^2
\end{equation}
where $
\Delta= ^2+a^2-2mr\,,
\rho^2 = r^2+a^2\cos^2\theta\,,  
\Sigma^2=  (r^2+a^2)^2-\Delta a^2\sin^2\theta\,, 
\psi^2= \Delta-a^2\sin^2\theta\,, 
\chi= 2amr$. 
Now we can use the standard procedure to calculate the corrections to the entropy of the Kerr
black hole from noncommutativity. Thus, if $\tilde{S}(r_+)$ is the entropy of the noncommutative Kerr
black hole, then we can write,
 \begin{eqnarray}
\tilde{S}(r_+)&=& S(r_+)-\frac{\pi\beta^2(E/E_p)}{8}\left\{8a^2
  +\left[\frac{(a^6+2a^4r_+^2-3a^2r_+^4+2r_+^6)}{ar_+(a^2+r_+^2)}\right]\right. \nonumber \\ &&\left. \times
\tan^{-1}{\left(\frac{a}{r_+}\right)}\right\}\,\nonumber \\ 
&=&  S(r_+)- \frac{\pi\gamma^2}{8}\left(\frac{ E^2}{ E_P^2 (E- E_P)^2}\right)^\eta \nonumber \\ && \times \left\{8a^2
  +\left[\frac{(a^6+2a^4r_+^2-3a^2r_+^4+2r_+^6)}{ar_+(a^2+r_+^2)}\right]
\tan^{-1}{\left(\frac{a}{r_+}\right)}\right\},
\end{eqnarray}
where $S(r_+)$ is the entropy for the commutative black hole.
Finally, we can write for the noncommutative Kerr black hole
\begin{eqnarray}
\frac{\partial r_+}{\partial M}&=&\frac{r_+}{r_+-M}\,,\\
\frac{\partial \tilde{S}}{\partial r_+}&=& 2\pi r_++\frac{\pi\beta^2(E/E_p)}{8}\Biggl
\{\left[\frac{(a^6+2a^4r_+^2-3a^2r_+^4+2r_+^6)}{r_+(a^2+r_+^2)^2}\right]\nonumber\\
&&+\left[\frac{\Phi}{ar_+^2(a^2+r_+^2)^2}
\right]\tan^{-1}{\left(\frac{a}{r_+}\right)}\Biggr\},
\end{eqnarray}
where $\Phi  = a^6(a^2+3r_+^2)-2a^4r_+^2(a^2-r_+^2)+3a^2r_+^4(3a^2+r_+^2)-2r_+^6(5a^2+3r_+^2)$.
So the temperature for the noncommutative Kerr black hole can be written as
\begin{eqnarray}
  T^{-1}&=& \left(\frac{\partial \tilde{S}}{\partial r_+}\right)\left(\frac{\partial r_+}{\partial M}\right)\,,
\nonumber \\ &=& \left[\frac{r_+}{r_+-M} \right] \left[ 2\pi r_++\frac{\pi\beta^2(E/E_p)}{8}\Biggl
\{\left[\frac{(a^6+2a^4r_+^2-3a^2r_+^4+2r_+^6)}{r_+(a^2+r_+^2)^2}\right]\right.\nonumber\\
&& \left. +\left[\frac{\Phi}{ar_+^2(a^2+r_+^2)^2}
\right]\tan^{-1}{\left(\frac{a}{r_+}\right)}\Biggr\} \right].
\nonumber \\ &=& 
\left[\frac{r_+}{r_+-M} \right] \left[ 2\pi r_++  \frac{\pi\gamma^2}{8}\left(\frac{ E^2}{ E_P^2 (E- E_P)^2}\right)^\eta \right. \nonumber \\ 
&& \left. \times \Biggl
\{\left[\frac{(a^6+2a^4r_+^2-3a^2r_+^4+2r_+^6)}{r_+(a^2+r_+^2)^2}\right]\right.\nonumber\\
&& \left. +\left[\frac{\Phi}{ar_+^2(a^2+r_+^2)^2}
\right]\tan^{-1}{\left(\frac{a}{r_+}\right)}\Biggr\} \right].
\end{eqnarray}
It may be noted that for this deformed solution, a
   a black remnant cannot form,  as these correction  also increase rather than reduce the temperature 
   of the black hole.   
  
Thus, we have obtained an expression for the corrections to the thermodynamics of Kerr black holes.
Here the metric, and the thermodynamics of the black hole depends on the energy of the particle in Hawking radiation. 
It is known that the energy of the particle in the Hawking radiation is different for an in-falling observer than an 
asymptotic observer \cite{hawk, hawk1, hawk2}. So, the geometry of the spacetime will also appear considerable different for the two 
observes.  It may be noted that the idea that the geometry of spacetime appear different for an  in-falling observer
and an asymptotic observer has been made for discussing the black hole complementarity   \cite{comp, comp1, comp2, comp4}. 
It has been observed that the gravity's rainbow also modifies the geometry of a black hole, and this 
modification has been used to address the black hole information paradox \cite{bhip, bhip1, bhip2}. 
It may be noted in the commutative limit these corrections vanish, and we obtain the original case back.

\section{Conclusion} 

In this paper, motivated by gravity's rainbow, we have proposed a new kind of energy dependent noncommutative geometry. 
We have studied  the
noncommutative Schwarzschild black holes and the noncommutative Kerr black holes, using this new energy dependent noncommutative geometry.
This was done by first using Weyl quantization
procedure~\cite{Wess,Weyl} to relate  the metric tensor on noncommutative
spacetime  with the metric on commutative spacetime. The relations thus derived were used specifically to
obtain expressions for the corrections to the Schwarzschild metric
and the   Kerr black metric due to noncommutativity. Furthermore,   we analyzed the thermodynamics of these
noncommutative black holes. So, explicit expressions for the
  corrected entropy and temperature of these black hole solutions were obtained. These corrections vanished
  when the noncommutative parameter is set to zero, and hence, we derive the original commutative results back.
It was observed that for  these deformed solutions,  
  black remnants cannot form. This is because these correction increase rather than reduce the temperature 
   of the black holes. 
   
  It may be noted that  this work take a different approach from the earlier works where
  the effect of noncommutativity has been  studied by    replacing
  point-like structures in the original theory by   objects which are smeared in
  spacetime \cite{a, b,c,d}.  
   The noncommutativity  mixes   ultraviolet and
infrared divergences \cite{az}. It also
 incorporates
non-locality  in a controllable way \cite{moff}. It has been studied
in the  context of string theory. This is because
it is known that the transverse coordinates of D-branes can be regarded
as matrices, and these matrices do not commute
\cite{Wi1}.  Noncommutativity also occurs in the context of M-theory and
 \cite{CoDoSc,KaOk}. In this analysis   
the  compactification on the noncommutative torus has been studied.
It has been demonstrated that   deforming the commutative torus to the noncommutative
torus corresponds to have a   constant background three form potential.
In fact, there are
  two commutative tori   associated with  a noncommutative torus,
one to its odd and one to its even cohomology, leading to two commuting
  actions
on the Teichmuller space. Noncommutative geometry has also been studied in the context
of  open strings ending on the D-branes.
 In this context    gauge theories on
 noncommutative tori will   appear as D-brane world volume theories \cite{DoHu}.
 The  D0-branes in type IIA  string theory with a background two form
 field have also been studied, and it has been observed that the
  background two form field modifies the
replacing  ordinary multiplication by a noncommutative product \cite{ChKr}.
 It has been demonstrated that there is a link between a
noncommutative gauge theory  and an ordinary gauge theory \cite{SeWi}.
 In fact, a relation between the noncommutative
  instantons and
the ordinary instantons  for  Yang-Mills theory has also been
observed
\cite{NeSc}.
 The relation between instantons on branes and the noncommutative Yang-Mills theory
 has also been observed \cite{Be}. In fact, the
  noncommutative instanton on the torus \cite{AsNeSc} and
the monopole in the noncommutative $U(2)$ Yang-Mills theory
 \cite{HaHaMo,HaHa} have also been studied.
It has been demonstrated that the  $U(1)$ effective action for branes is the
 the Dirac-Born-Infeld action
\cite{Ts}, and the
BPS condition of the ordinary Dirac-Born-Infeld action and
a noncommutative action are equivalent in a limit $\alpha' \rightarrow 0$ \cite{SeWi}.
It may be noted that   string
theory  also can give rise to noncommutative gravity  \cite{ng}. This is done   by studding
the next to the leading order terms in the Seiberg-Witten limit for 
the dynamics of closed strings, in the presence of a constant
two form field. Thus,  the
  gravitational action induced by the bosonic string theory on a space-filling D-brane
with a constant magnetic field have been studied in the low energy limit. The
   induced terms for
the interaction vertex of three gravitons on the brane have thus been obtained.
  It has also been observed that the  noncommutative
deformations of gravity can lead  to a complex metric and in this case the tangent space
groups is larger than the Lorentz group \cite{az12}.
Noncommutativity has also been studied by twisting the
  diffeomorphism invariance of the general
relativity \cite{W}. It may be noted that the deformed algebra corresponding to commutative
deformation has been used to
construct a covariant tensor calculus for metric, covariant derivatives, curvature and torsion \cite{W}.
It would be interesting to analyse all these structures using this new form of noncommutativity, which is motivated 
from gravity's rainbow.

  The effect of having an energy dependent 
  metric on the thermodynamics of various black objects has already been studied using gravity's rainbow \cite{5A, 6B, 2A, ga}, 
  it  would be interesting to analyse such an energy dependence using the approach developed in this paper. Furthermore,
  the thermodynamics of various interesting   black objects has been studied 
  \cite{ring, ring2, ring4, ring5, ring7, ring8}, and it would be interesting to analyse the effect of energy dependent noncommutativity 
  on the thermodynamics of such black objects.  It is expected that the black hole remnants will not form for all the black objects due to this energy 
  dependent noncommutative deformation of spacetime. It is expected that this deformation will increase the temperature rather than 
  decrease it for all black objects. However, it would be interesting to demonstrate it explicitly for various different black hole solutions. 
  In higher dimensions, interesting solutions to the general relativity exist which have interesting topologies.
  It is possible for  black rings and black saturns to exist in higher dimensions. It will be interesting to
  analyze the thermodynamics of such solutions using noncommutative formalism. It has also been demonstrated that
  noncommutativity leads to the existence of a minimum length in spacetime. The black hole thermodynamics for
  minimum length have been analyzed \cite{gup, 0gup, 1gup, gup1}. It will be interesting to analyze a possible
  link between the thermodynamics
  of black holes with minimum length and noncommutative black holes. It would also be interesting to generalize this analyse to 
  some energy dependent version of such a deformation.  It may be noted that this noncommutative deformation will deform the 
  Heisenberg algebra, and it is known that a deformation of the Heisenberg algebra is related to generalized uncertainty principle, 
  which in turn is related to modified dispersion relation \cite{mdr12, mdr14}. It would be interesting to investigate further the
  relation between this energy dependent noncommutativity and modified dispersion relation in flat spacetime.

\end{document}